# The nature of transmission resonances in plasmonic metallic gratings

G. D'Aguanno,[1,2,*] N. Mattiucci,[1,2] M. J. Bloemer,[2] D. de Ceglia[1,2], M.A. Vincenti[1,2], and A. Alù[3]

[1] AEgis Tech. Microsystems/Nanotechnology, 410 Jan Davis Drive, Huntsville AL 35806, USA

[2] Dept. of the Army, Charles M.Bowden Facility , Bldg 7804, Research Development and Engineering Command , Redstone Arsenal, AL 35898,USA

[3] Department of Electrical and Computer Engineering, University of Texas at Austin, Austin, TX 78712, USA

*Corresponding author: giuseppe.daguanno@us.army.mil

**Abstract:** Using the Fourier modal method (FMM) we report our analysis of the transmission resonances of a plasmonic grating with sub-wavelength period and extremely narrow slits for wavelengths of the incoming, transverse magnetic (TM)-polarized, radiation ranging from 240nm to 1500nm and incident angles from $0^0$ to $90^0$. In particular, we study the case of a silver grating placed in vacuo. Consistent with previous studies on the topic, we highlight that the main mechanism for extraordinary transmission is a TM-Fabry-Perot (FP) branch supported by waveguide modes inside each slit. The TM-FP branch may also interact with surface plasmons (SPs) at the air/Ag interface through the reciprocal lattice vectors of the grating, for periods comparable with the incoming wavelength. When the TM-FP branch crosses a SP branch, a band gap is formed along the line of the SP dispersion. The gap has a Fano-Feshbach resonance at the low frequency band edge and a ridge resonance with extremely long lifetime at the high frequency band edge. We discuss the nature of these dispersion features, and in particular we describe the ridge resonance in the framework of guided-mode resonances (GMRs). In addition, we elucidate the connection of the coupling between the TM-FP branch and SPs within the Rayleigh condition. We also study the peculiar characteristics of the field localization and the energy transport in two topical examples.

# 1. Introduction

Scattering of the electromagnetic radiation from metallic gratings [1] has attracted much attention since the beginning of the last century, when Wood [2] noted an uneven distribution of the diffraction orders reflected from them under TM polarization of the incident light (i.e. the H-field parallel to the grooves of the grating). The phenomenon was later explained by Ritchie et al. [3] in terms of an interaction between the incoming photon and a surface-plasmon (SP) resonance at the grating surface. The interested reader may consult Ref. [4] for an account of the influence of SP resonances on the diffraction anomalies of metallic gratings. The subject took new life approximately one decade ago when Ebbesen and coworkers [5] experimentally verified enhanced transmission in the optical regime through an array of strongly sub-wavelength holes carved in a metallic screen under TM light polarization. The topic of the extraordinary transmission through these kinds of structures has been the subject of an intense theoretical and experimental investigation over the past decade, an account of which can be found in Refs. [6-7]. The results of Ref. [5] sparked a renewed interest in the scattering from sub-wavelength metallic gratings which can be considered in some ways the one-dimensional version of the holey metallic screen studied in Ref. [5]. In Ref. [8] the authors theoretically showed the existence of transmission resonances at wavelengths larger than the period of the grating for extremely narrow slits. In particular, they enucleated two different resonant mechanisms for the energy transport from the input to the output surface of the grating: a) the excitation and coupling of SP resonances at the input and output surfaces of the grating and b) the coupling of light through the waveguide resonances located in the slits. In particular, as it is well known [9], the fundamental TEM mode in a planar parallel-plate waveguide made of a perfect conductor has no cut-off and, therefore, a guided mode exists even for strongly sub-wavelength slits, a mechanism that allows wave transmission through ultranarrow slits. Such propagation, instead, is not supported in the case of a rectangular or cylindrical conducting waveguide, for which a cut-off dimension exists. This fact implies that waveguide modes cannot be invoked to justify the extraordinary transmission in the case of holey screen, as studied in Ref.[5]. There has been some debate on whether the coupling with waveguide modes in the slits is sufficient to explain the extraordinary transmission in sub-wavelength metallic gratings [10] or the SP resonances are in any case essential to it [7,11], and we discuss this issue in the following. It should be also stressed that at optical frequencies metals lose some of their conductive properties, affecting the TEM mode in parallel-plate slits, which is transformed into a plasmonic TM mode with slow-wave properties.

Analogous transmission properties may be verified based on this dominant mode, which also has no cut-off. At optical frequencies the propagation loss of these TM waveguide modes may become quite large for long propagation distances, but the length of the waveguide slits considered here is on the order of 100 nm, and significant amounts of energy may be expected to be transmitted through the slits via the waveguide modes. In the case where there are no waveguide modes, even evanescent tunneling through the apertures may be a mechanism for energy transport in specific configurations.

In this paper, we verify that the main mechanism for the extraordinary transmission of TM waves through 1-D ultra-narrow slit arrays in a metallic grating is mainly based on a Fabry-Perot (FP) branch which depends on waveguide modes inside each slit. We also discussed how the TM-FP branch in several circumstances can be coupled with the surface plasmons (SPs) supported at the input and output surfaces of the grating through the reciprocal lattice vectors of the grating. When the TM-FP branch interacts with these SPs, a band gap is formed along the line of the SP dispersion. The gap has a Fano-Feshbach resonance at the low frequency band edge and a ridge resonance with extremely long lifetime at the high frequency band edge, which may be interpreted in the framework of guided-mode resonances (GMRs). We elucidate the connection of the coupling between the TM-FP branch and SPs with the Rayleigh condition and the Wood's anomaly. The paper is organized as follows: in Section 2 we briefly describe the geometry of interest and the numerical tool we use for the calculations. In Section 3 we analyze in detail the TM-FP branch. In Section 4 we describe the coupling of the TM-FP branch with the SPs and elucidate the connection with the Fano-Feshbach resonance and the ridge resonance. In Section 5 we discuss the Rayleigh condition. In Section 6 we analyze the peculiar features of the field localization and the energy transport in two topical cases. Finally in Section 7 we give our conclusions.

## 2. The model

In our analysis we use the Fourier Modal Method (FMM) [12] to study the transmission resonances from a silver grating with strongly sub-wavelength slits for a range of the incoming radiation which encompasses the extreme UV, visible and near-IR part of the electromagnetic spectrum, namely from 240nm to 1500nm, and incident angles from $0^0$ to $90^0$. The geometry studied is sketched in Fig.1. The inset of Fig. 1 shows the dispersion of silver as reported in Palik's handbook [13]. The use of the FMM allows us to take directly the values of the electric permittivity of silver from measured data available in literature [13], with no fitting procedure and no auxiliary equations accounting for the Drude-Lorentz dispersion model as would be the case if one uses other numerical techniques such as finite-difference time-domain (FDTD) [14], for example. Note that the oscillations in the real part of the permittivity in the long wavelength region result from the overlap of data points obtained from different experimental groups [13].

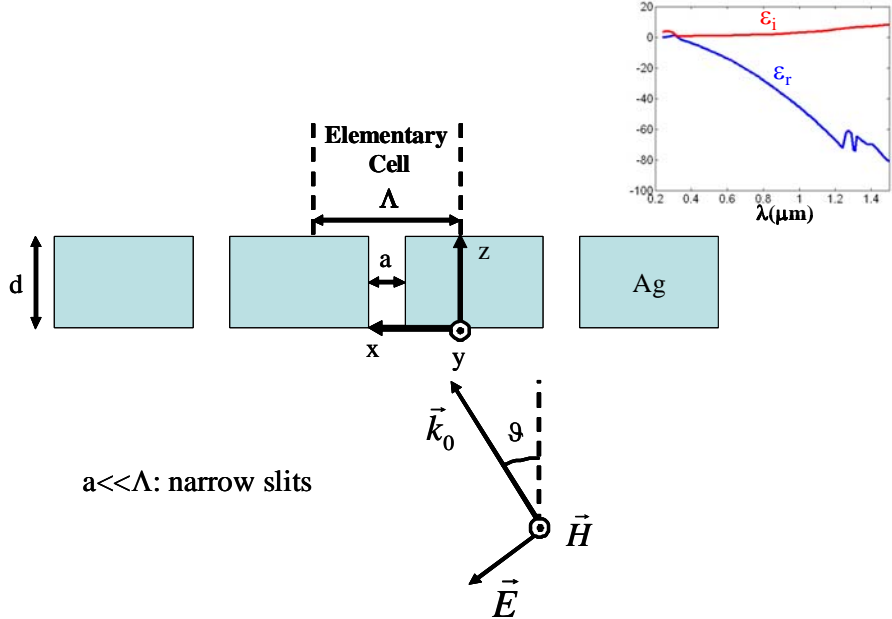

Fig. 1. Metallic grating made of silver with grating thickness $d$, slit aperture $a$ and grating period $\Lambda$. The incident field is a plane, monochromatic, TM-polarized wave where $k_0=2\pi/\lambda$ is the wave-vector, $\lambda$ is wavelength and $\vartheta$ is the incident angle. In this case we consider air as the incident and the output medium. The Cartesian right-handed system (x,y,z) has the z coordinate in the propagation direction and the x coordinate along the periodicity of the grating. The y coordinate is the direction along which the magnetic field H is polarized. The input grating surface is located at z=0 and the output surface at z=d. Inset: Real ($\varepsilon_r$) and imaginary ($\varepsilon_i$) part of the electric permittivity of silver from data reported in Ref.[13].

The FMM has been largely used in the past to describe the scattering from diffraction gratings [15]. In its simplest form the method is based on the Fourier expansion of the grating profile and the resolution of a coupled system of damped, harmonic-oscillator-type equations [15]. Despite its simplicity and straightforwardness, the method in the past suffered from two serious problems which prevented its full applicability especially in the plasmonic regime. The first problem was the poor convergence for TM polarization and the second was the numerical instabilities which may arise when matching the solution inside the grating with the solution in the incident and output medium due to the presence of evanescent modes. The first problem was resolved almost fifteen years ago in a series of seminal papers where it was made clear that the slow convergence was due to the way in which the Fourier transform of the product of two discontinuous functions is handled [16-18] and suggested a way to manage the expansions in the proper manner. Further refinements of the technique for particular cases have been subject of investigations until very recently [19]. The numerical instabilities due to the presence of evanescent modes can be cured in two different ways. The first way is by resorting to the scattering-matrix (S-matrix) approach [20] instead of the classical transfer matrix approach or by resorting to the R-matrix approach which is a slightly different version of the S-matrix approach [20]. The S-matrix and R-matrix approaches are both recursive methods that achieve unconditional numerical stability [20] and require moderate computer memory. The second, much simpler method, nevertheless still achieving numerical stability, is the one laid out in Ref. [21] where the boundary value problem is resolved simultaneously at all the grating interfaces. This second method requires more computer memory but, on the other hand, has the advantage of a much easier and intuitive implementation.

In our version of the FMM we follow the classical recipe suggested in Refs. [16-18] to cure the poor convergence of the TM polarization and we solve simultaneously the boundary conditions at the input and output surfaces of the grating according to Ref. [21]. The transmitted power at the output surface of the grating is calculated by using the z-component of the Poynting vector ($S_z$), namely:

$$T(\lambda, \vartheta) = \frac{\int_0^\Lambda S_z(x, z = d, \lambda, \vartheta) dx}{\int_0^\Lambda S_z^{input}(x, z = 0, \lambda, \vartheta) dx} \quad . \quad (1)$$

In other words, T is calculated by the ratio of the power transmitted by the elementary cell of the grating at the output surface located at *z=d* divided the power incident onto the elementary cell of the grating located at *z=0* (see Fig.1). Using the orthogonality condition of the diffraction orders [22], Eq.(1) can be recast for TM polarization in the following form:

$$T(\lambda, \vartheta) = \frac{n_{inc}}{k_0 n_{out}^2 \cos \vartheta} \sum_m |t_m|^2 \text{Re}\left[\sqrt{n_{out}^2 k_0^2 - \alpha_m^2}\right] \quad , \quad (2)$$

where $k_0 = 2\pi/\lambda$ is the vacuum wave-vector, $n_{inc}$ and $n_{out}$ are respectively the refractive index of the input and output medium ($n_{in} = n_{out} = 1$ in our case), $\vartheta$ is the incident angle of the incoming wave on the grating, $t_m$ is the complex transmission coefficient of the *m-th* diffracted order, Re indicates the real part, and finally $\alpha_m$ is the generalized transverse wave-vector:

$$\alpha_m = k_0 n_{inc} \sin\vartheta + \frac{2m\pi}{\Lambda} \quad m=0,\pm1,\pm2,... \quad . \tag{3}$$

We have written Maxwell's equations in MKSA, non-dimensional units, i.e. we write the equations as in the standard MKSA units, but then we take $\varepsilon_0=\mu_0=c=1$ where $\varepsilon_0$ is the vacuum electric permittivity, $\mu_0$ is the vacuum magnetic permeability, and $c$ is the speed of light in vacuo. Finally, the magnetic field incident onto the sample is considered of unitary amplitude.

## 3. The TM-FP Branch: the Role of Plasmon Modes in the Metal-Insulator-Metal (MIM) Waveguide

Let us start by investigating in detail TM-FP branch which is the main protagonist of the extraordinary transmission through the geometry of Fig. 1. At this regard in Fig.2 (a) we plot the transmission T in the ($\lambda$,d) plane for normal incidence in the case of the Ag grating with period $\Lambda$=320nm and slits aperture a=32nm. In this case we let its thickness vary between d=10nm and d=400nm. Fig.2(a) clearly puts into evidence the Fabry-Perot nature of the branch, in fact by varying the thickness of the grating we can see several transmission branches located respectively at d~$\lambda_g$/2, d~$\lambda_g$, d~3$\lambda_g$/2 where $\lambda_g$=$\lambda$/n$_{eff}$ is the wavelength of the guided mode with effective index n$_{eff}$, as clarified below. The white dashed lines, reported for comparison, represent, respectively, the position of the first, second and third transmission branch calculated by finding the maxima of the square modulus of the complex transmission function of a FP-etalon interferometer with same effective index placed in vacuo for TM polarization of the incident light [23]:

$$t_{TM}(k_x,\omega) = \frac{2}{2\cos\left(\hat{n}_{eff}\sqrt{k_0^2 - \frac{k_x^2}{\hat{n}_{eff}^2}}d\right) - i\left(\frac{\hat{n}_{eff}\sqrt{k_0^2 - \frac{k_x^2}{\hat{n}_{eff}^2}}}{\hat{\varepsilon}_{eff}\sqrt{k_0^2 - k_x^2}} + \frac{\hat{\varepsilon}_{eff}\sqrt{k_0^2 - k_x^2}}{\hat{n}_{eff}\sqrt{k_0^2 - \frac{k_x^2}{\hat{n}_{eff}^2}}}\right)\sin\left(\hat{n}_{eff}\sqrt{k_0^2 - \frac{k_x^2}{\hat{n}_{eff}^2}}d\right)} \tag{4}$$

Note that in Eq.(4) the ω dependence is in the vacuum wavevector $k_0=\omega/c=2\pi/\lambda$. The FP-etalon has thickness $d$ and complex effective index $\hat{n}_{eff} = n_{eff} + iK_{eff} = \sqrt{\hat{\varepsilon}_{eff}}$, where $n_{eff}$ and $K_{eff}$ are respectively the effective index and the extinction coefficient for the fundamental TM guided mode of a subwavelength metal-insulator-metal (MIM) planar waveguide [24] made in our case by Ag/air(32nm)/Ag. Fig. 2(b) shows the effective index of the fundamental mode for the Ag/air/Ag waveguide, varying the thickness of the air core. The calculation has been performed using a Newton-Rapson procedure as in Ref. [25]. Note that in the limit of a thick air core (i.e. the thickness of the core is larger than the skin depth of the plasmon in air) the dispersion of the guided mode tends toward the dispersion of the single interface air/Ag SP (last curve on the bottom of the figure ) in agreement with the results reported in Ref. [24]. The effective index of the SP is given by [4]:

$$n_{eff,SP} = \mathrm{Re}\sqrt{\frac{\hat{\varepsilon}_{Ag}}{1+\hat{\varepsilon}_{Ag}}} \quad , \tag{5}$$

where $\hat{\varepsilon}_{Ag}$ is the complex permittivity of silver taken from Ref. [13].

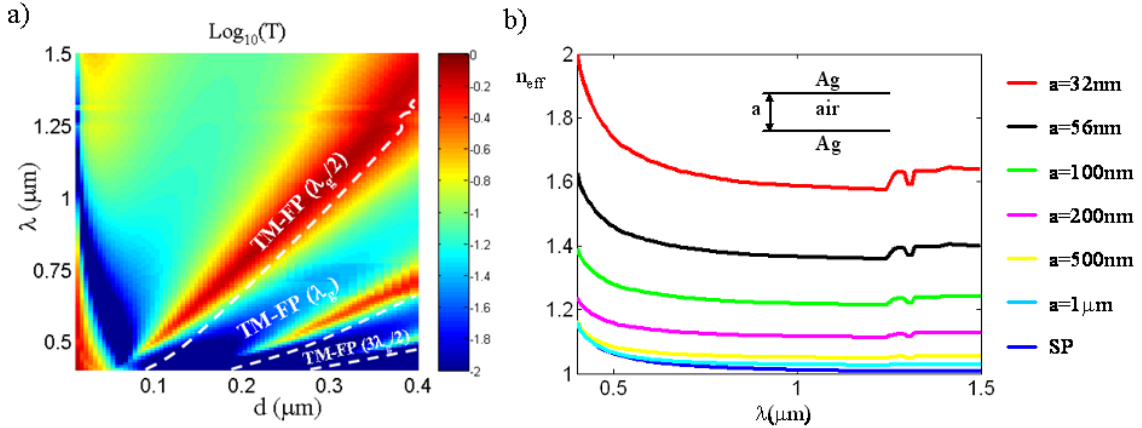

Fig.2. a) $\mathrm{Log}_{10}(T)$ vs. $\lambda$ (incident wavelength) and $d$ (grating thickness) for normal incidence. The grating parameters are: $\Lambda$=320nm and a=32nm. The various TM-FP branches are put into evidence. The dashed white lines represent the various FP branches calculated using the FP-etalon formula with the complex effective index of the guided mode. b) Effective index $n_{eff}$ vs. $\lambda$ for the fundamental TM mode of a planar waveguide Ag/air/Ag for different air core thicknesses. The various dimensions of the air core for the different curves from the top to the bottom are reported at the right side of the figure. The last curve on the bottom represents instead the effective index of the SP at a single air/Ag interface. Inset: schematic representation of the MIM waveguide.

From Fig. 2(a) and 2(b) several conclusions may be drawn: a) The position of the FP-branches is, everything considered, well approximated by the position of the branches in an equivalent FP-etalon of thickness $d$ and complex effective index $\hat{n}_{eff} = n_{eff} + iK_{eff}$. b) The $\lambda_g/2$ branch, i.e. the first

FP branch, is pushed toward ~$\lambda/4$ by the effective index of the waveguide mode (for a=32nm we have $n_{eff}$~2 in the visible range). The same effect can be found for example in the nano-antennas where the resonant frequency is shifted at $\lambda_{eff}/2$ where $\lambda_{eff}$ is the effective wavelength of the plasmonic nano-dipole which is shorter than the free space wavelength $\lambda$ [26-27]. This is a fundamental distinction with respect to the microwave or THz region. If we had studied the same geometry (i.e. with the dimensions scaled according to the wavelength) in the THz or microwave range we would have found that the first FP resonance would have been located at ~$\lambda/2$. The fact that we find the first FP branch at ~$\lambda/4$ ($\lambda_g/2$) is a typical signature of the plasmonic regime. c) The dispersion of the guided mode clearly shows its plasmonic nature, given the fact that it tends toward the dispersion of the SP of the single air/Ag interface by increasing the thickness of the air core.

It should be therefore clear that the TM-FP branch in the optical regime is actually ruled by a guided mode in the z-direction that is plasmonic in nature, as the modes studied in Ref. [24] for the MIM planar waveguide. While at lower frequencies (say THz and below) the differences between classical TEM-modes in a perfect conductor and these plasmonic modes become much less evident, in the optical regime, that is the range of interest of the present work, the plasmonic nature of these modes is preponderant. In Fig. 3 we show the transmission T in the ($\omega$,$k_x$) plane for the Ag grating with thickness d=400nm, period $\Lambda$=256nm and slits aperture a=32nm. In the figure it is also reported (black-dashed line) the dispersion $\omega_{FP}(k_x)$ of the equivalent FP-etalon which is calculated by finding the transmission maxima of $|t_{TM}(\omega,k_x)|^2$ where $t_{TM}(\omega,k_x)$ is given by Eq.(4).

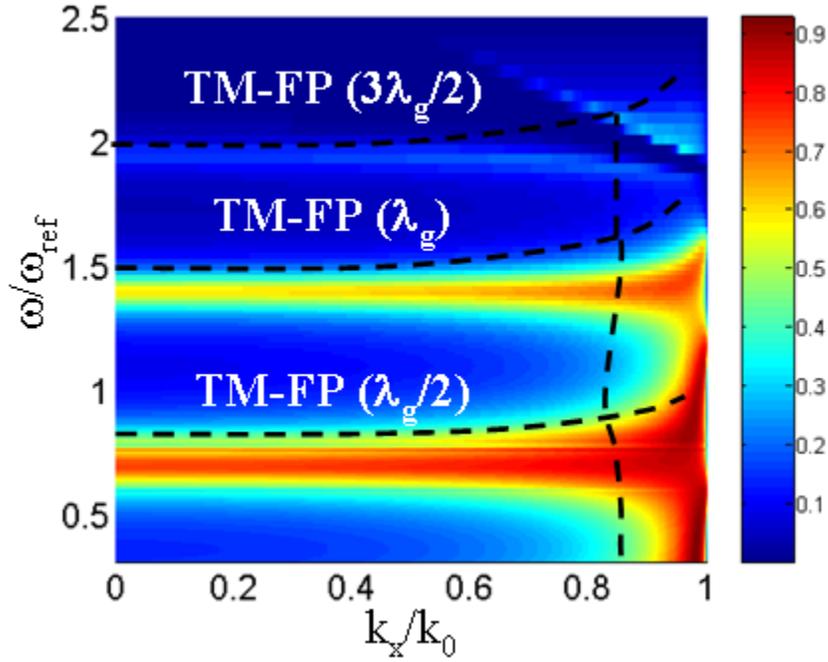

Fig. 3. T vs. $\omega/\omega_{ref}$ and $k_x/k_0=\sin(\vartheta)$. $\omega_{ref}$ is a reference frequency which corresponds to $\lambda=1\mu m$. The $\omega/\omega_{ref}$ scale goes from $\omega/\omega_{ref}=0.667$ ($\lambda=1.5\mu m$) to $\omega/\omega_{ref}=2.5$ ($\lambda=400nm$). The grating parameters are: $\Lambda=256nm$, $d=400nm$, $a=32nm$. The dashed line corresponds to the dispersion of the equivalent FP-etalon for which the nearly vertical dashed line is related to the classic Brewster condition of a dielectric slab..

The figure shows that the salient characteristics of the TM-FP ($\lambda_g/2$) branch and the TM-FP ($\lambda_g$) branch are described quite satisfactorily by the dispersion of the equivalent FP-etalon. Clearly, this is only a qualitative analogy, which captures the position of the maxima of transmission related to the Fabry-Perot resonances of the grating, but it provides physical insights into the optical behavior of the grating in this regime. In the figure it is also visible the TM-FP ($3\lambda_g/2$) branch that is in this case split by the dispersion of the transverse SPs which make their appearance in the propagation region thanks to their phase matching with the reciprocal lattice vectors of the grating. The detailed description of the coupling between the TM-FP branch and the SPs at the input and output surfaces will be the subject of the next Section. For the time being let us say that Eq.(4) gives just a rough estimation of the position of the FP resonances, in fact it does not take into account the grating periodicity and the possible interactions between the multiple slits, all factors that affect the actual position of the resonances. Still, even within this crude approximation, the main features of the FP branch are recaptured in an acceptable way. It is relevant to stress that the position of the maxima predicted by Eq. (4) are slightly shifted compared to the calculated transmission resonances in Fig. 2-3. This is an expected phenomenon even in conducting slits, associated with the fact that the phase of the reflection coefficient at the slit apertures is in practice not purely real, as assumed in Eq. (4) [10], even in the lossless limit. An improved model for the effective properties of the grating in this long-wavelength regime will be the subject of a future more detailed investigation that we are currently performing. We would like to close this Section by remarking the fact that the Fabry-Perot nature of these transmission resonances has been widely studied in the past [11, 28-35]. In particular the first clear claim about the transmission resonances at normal incidence following the Fabry-Perot resonant condition $d=m\lambda_g/2$ is reported in Ref.[28]. An up to date review on the subject can be found in Ref.[36]. Our approach here is characterized by its extreme simplicity, in fact Eq.(4) is the standard Fabry-Perot-etalon formula for generic angular incidence and TM polarization where the refractive index of the Fabry-Perot-etalon is nothing more than the complex effective index the fundamental guided mode of the MIM waveguide. Moreover we have clearly demonstrated the plasmonic nature of these Fabry-Perot transmission resonances in the optical regime.

## 4. The coupling between the TM-FP Branch and SPs at the input/output interfaces

When the grating period becomes comparable to the impinging wavelength, the dominant TM-FP resonant branch highlighted above may couple with the SPs supported by the input and output interfaces of the grating. To the end of modeling this coupling, we write the energy-momentum conservation of the incoming photons for both processes:

$$\omega_{ph}(k_x) = \omega_{FP}(k_x) \quad , \quad (6.a)$$

$$\omega_{ph}(k_x) = \Omega(|k_{SP} \pm G_m|) \ m = 0,1,2,... \quad . \quad (6.b)$$

In Eq.(6.a) $\omega_{ph}$ is the energy of the incoming photon, $k_x = k_0 sin(\vartheta)$ its transverse momentum, $\omega_{FP}(k_x)$ the dispersion of the equivalent FP-etalon. In Eq.(6.b) $\Omega$ is the plasmon energy, $k_{SP} = k_0 \, \text{Re} \sqrt{\hat{\varepsilon}_{Ag}/(1+\hat{\varepsilon}_{Ag})}$ the momentum of the single interface SP for the homogeneous (no patterning) air/silver interface [3,4] and $G_m = 2m\pi/\Lambda$ is the *m-th* reciprocal lattice vector of the grating. Eq. (6.b) refers to the transverse SP, i.e. the SP that is guided along the x-axis at the air/Ag interface. In our case we deal with thick gratings (d≥150nm) and therefore the use of the SP dispersion of the single homogeneous interface is a very good approximation of the SP modes supported by the layer. Actually the plasmonic dispersion relation of a thin metal layer automatically converges toward the dispersion of the single interface plasmon for layer thickness greater than 50nm, as shown in, e.g., Ref. [37]. In Fig. 4 we show two typical examples of the coupling between the TM-FP branch and the dispersion of the SPs.

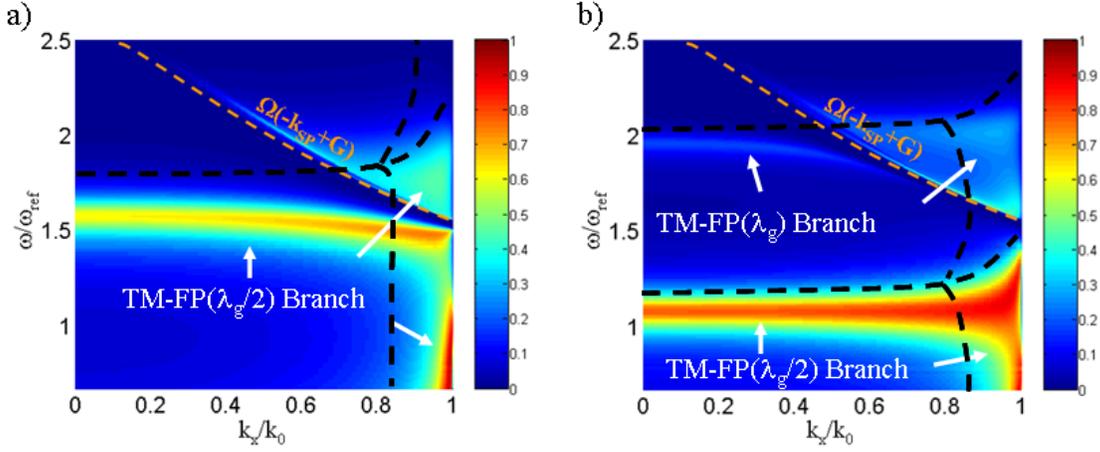

Fig.4. T vs. $\omega/\omega_{ref}$ and $k_x/k_0$. The grating parameters common to both figures are: $\Lambda$=320nm, a=32nm. The thickness is respectively a) d=150nm and b) d=250nm. In both figures the thin-dashed line represents the dispersion of the SPs phase matched with the first reciprocal lattice vector of the grating: $\Omega(-k_{SP}+G)$. The thick-dashed line is the dispersion of the equivalent FP-etalon: $\omega_{FP}(k_x)$.

In Figs. 4 the thin-dashed line reports the dispersion of the transverse SP $\Omega(-k_{SP}+G)$ matched with the first reciprocal lattice vector and the black thick-dashed line is the dispersion of the TM-FP etalon $\omega_{FP}(k_x)$. It is important to realize that when the TM-FP branch and the SP dispersion cross paths, it is the SP that has the 'right of way'. In fact, the FP branch is split into two sub-branches separated by the line of the SP dispersion which in turn creates a band-gap between the two sub-branches. The lower sub-branch has to bend following closely the line of the plasmonic dispersion. In Fig.4(a) this happens for the TM-FP ($\lambda_g/2$), while in Fig.4(b) the same happens for the TM-FP ($\lambda_g$) branch. In Figs. 5, we show a magnification of Fig.4(a) near the SP dispersion line.

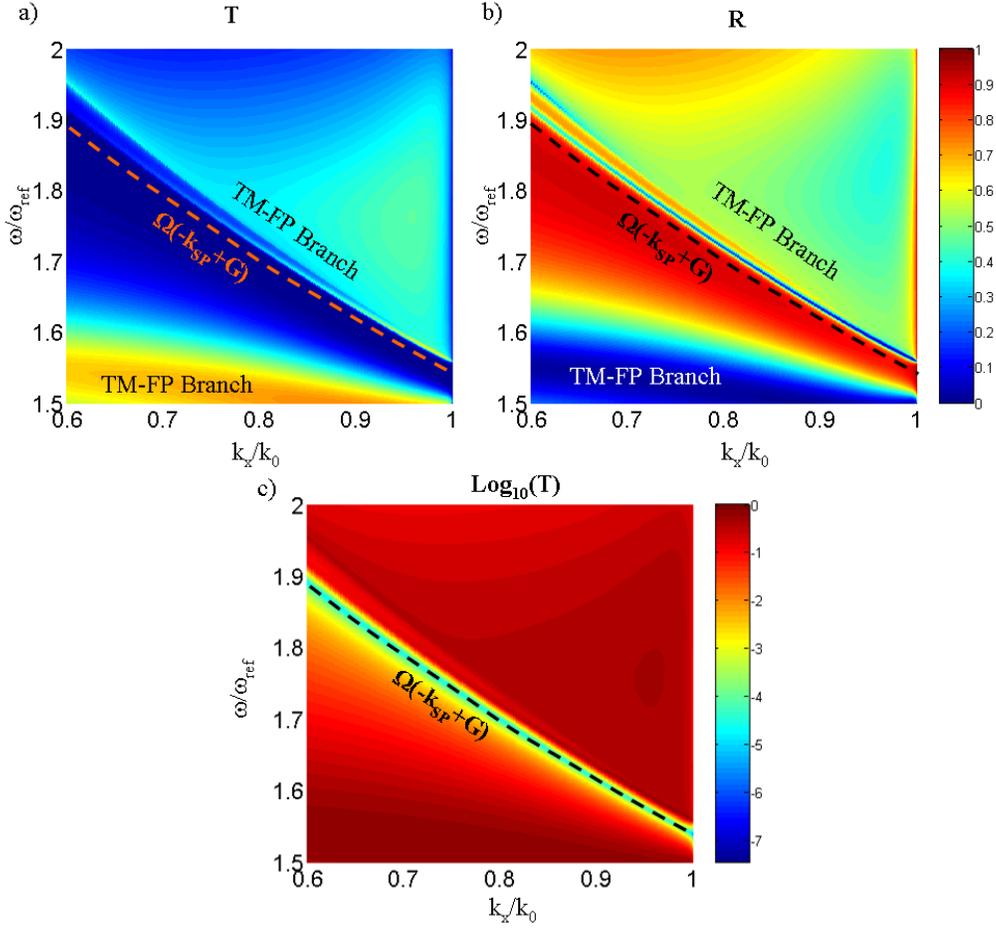

Fig. 5 (a) T vs. vs. $\omega/\omega_{ref}$ and $k_x/k_0$ for the case of Fig.4(a). This is a magnification near the SP dispersion line. (b) R (reflection) vs. $\omega/\omega_{ref}$ and $k_x/k_0$. (c) $Log_{10}(T)$ vs. vs. $\omega/\omega_{ref}$ and $k_x/k_0$. Note that the SP dispersion line (dashed line) follows exactly the transmission minima.

Fig. 5(a) clearly shows that the SP dispersion line shapes the gap formed between the two TM-FP sub-branches. In Fig. 5(b) we show in the same zone the reflection $R(\omega,k)$. The figure shows that the reflection peaks are almost coincident with the SP dispersion line. Note also that these peaks of reflection are effectively the Wood's anomalies of this geometry [2-3]. In Fig. 5(c) we also show $Log_{10}(T)$ from which it is clear that the minima of transmission exactly follow the SP dispersion line. The concept that the transmission minima follow the line of the SP dispersion will be instrumental to discuss the Rayleigh condition in the next Section.

For the time being let us analyze in detail in Fig. 6 the band-edge transmission resonances formed when the TM-FP branch is split by the coupling with the SP branch. In particular, we show in this figure a section of Fig. 4(a) taken for $k_x/k_0=\sin(80^0)$ in the frequency range around the SP frequency $\Omega(-k_{SP}+G=k_0\sin(80^0))$.

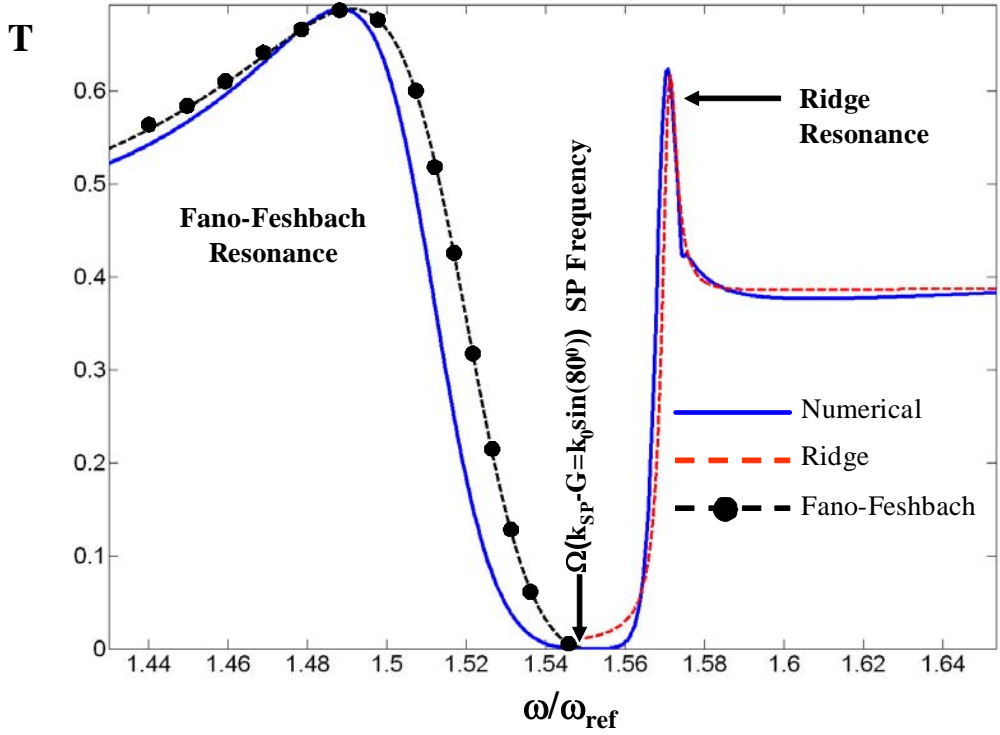

Fig. 6. T vs. $\omega/\omega_{ref}$ at $k_x/k_0=\sin(80^0)$ for the case of Fig.4 (a). The dispersion of the SP that crosses the TM-FP branch forms a band gap. The low frequency band-edge resonance has the form of a Fano-Feshbach resonance, while the high frequency band edge resonance has the form of a "ridge" resonance with extremely long life-time. Numerical calculation (continuous line), comparison with the Fano-Feshbach (F-F) resonance (dashed-dotted line) and with the ridge resonance (dashed line). The F-F resonance and the ridge resonance are calculated respectively according to Eqs(7-8). $\Omega=1.5513$ is the SP frequency at $k_{SP}-G=k_0\sin(80^0)$. From the numerical calculation we can extract the parameters to fit the resonances which are:, $\Omega_{FF}=1.4885$ (resonance frequency for the F-F), $\Gamma_{FF}=0.06$ (lifetime of the F-F), $\Omega_{ridge}=1.5708$ (resonance frequency of the ridge corresponding to $\lambda=637$nm), $\Gamma_{Ridge}=0.0045$ (lifetime of the ridge) corresponding to $\Delta\lambda\sim1$nm.

From the numerical calculation, we can extract the parameters that we need in order to write the various types of resonances as detailed in the figure caption. Once these parameters are computed, the Fano-Feshbach resonance normalized to the peak of emission ($T_{MAX}$) of the actual resonance can be calculated in the usual way [38-39]:

$$T_{F-F}(\omega;\Omega,\Omega_{FF},\Gamma_{FF},T_{MAX})=\frac{T_{MAX}}{1+q^2}\frac{(\varepsilon+q)^2}{\varepsilon^2+1} \quad , \qquad (7)$$

where, in this case $\varepsilon=2(\omega-\omega_F)/\Gamma_{FF}$, $q=2(\omega_F-\Omega)/\Gamma_{FF}$ and $\omega_F=(\Omega\Omega_{FF}+\Gamma^2_{FF}/4)^{1/2}$, here all the frequencies are intended in units of $\omega_{ref}$. The ridge resonance is instead described by the following expression:

$$T_{Ridge}(\omega;\Omega_{Ridge},\Gamma_{Ridge},T_{MAX})=T^2_{MAX}\left[T_{B-W}(\omega;\Omega_{Ridge},\Gamma_{Ridge},g=1)+\frac{1}{\pi}\arctan\left[\frac{4}{\Gamma_{Ridge}}(\omega-\Omega_{Ridge})\right]+\frac{1}{2}\right], \quad (8)$$

where $T_{B-W}$ is the Breit-Wigner ("Lorentzian") resonance defined as:

$$T_{B-W}(\omega;\Omega_{Ridge},\Gamma_{Ridge},g)=g\frac{(\Gamma_{Ridge}/2)^2}{(\omega-\Omega_{Ridge})^2+(\Gamma_{Ridge}/2)^2}. \quad (9)$$

Although outside the scope of the present work, the exploration and the exploitation of these extremely narrow ridge-resonances for the enhancement of nonlinear optical phenomena will surely be a fertile ground of research. We would like to caution the reader that our approach to the description of the transmission resonances in terms of a Fano-Feshbach line and a ridge resonance as described by Eqs. (7-9) is a qualitative description of the coupling between these two resonant mechanisms, with no pretension of rigor. Our scope here is to highlight the main "ingredients" that lay behind the physics of the enhanced transmission. A rigorous analytical theory is beyond the scope of the present work.

Nevertheless, our analysis highlights that the extraordinary transmission is deeply rooted in the physics of scattering, and somehow the "universality" of this phenomenon should be manifest. In this regard, more effective tools for its description and interpretation could be possibly taken directly from the field of atomic and nuclear physics, see, e.g., [40]. It is also interesting to remark the close similarity of the band gap structure shown in Fig. 6 with the band gap structure that is created in a simple 1-D multilayered structure as the one studied in, e.g., Ref. [41].

Let us elaborate further on the nature of this ridge resonance that appears at the high frequency band edge near the line of the dispersion of the plasmon coupled with the first reciprocal lattice vector. The ridge resonance may also be described in the framework of the 'guided-mode resonances' (GMRs) found in single slab or multi-layer waveguides with a grating on it. GMRs are leaky modes with extremely sharp transmission/reflection resonances which are originated by the coupling of the true guided modes of the waveguide with the reciprocal lattice vectors of the grating. GMRs have been widely studied in the past [42-45] and, obviously, they do not necessarily need plasmonic guided modes or surface waves to be generated. Indeed conventional guided modes coupled with the reciprocal lattice vector of the grating can be used to generate these sharp transmission/reflection resonances. In this regard, in Fig. 7 we show the transmission resonances at normal incidence for both TM and transverse-electric (TE) polarization in the case of a dielectric, non-dispersive grating of refractive index n=4.13 and thickness d=96nm. The grating period is Λ=320nm and the slit aperture is a=32nm. The thickness has been chosen so that the uniform layer supports at normal incidence a Fabry-Perot-etalon anti-resonance for $d=3\lambda/(4n)=96$nm at λ=532nm. For such a structure the impinging plane wave can indeed couple with the transverse guided TE and TM modes of the uniform layer and with its longitudinal Fabry-Perot resonances through the reciprocal lattice vectors of the grating, as verified comparing the transmission curves of the grating to the case of the simple etalon of same thickness (Fig. 7).

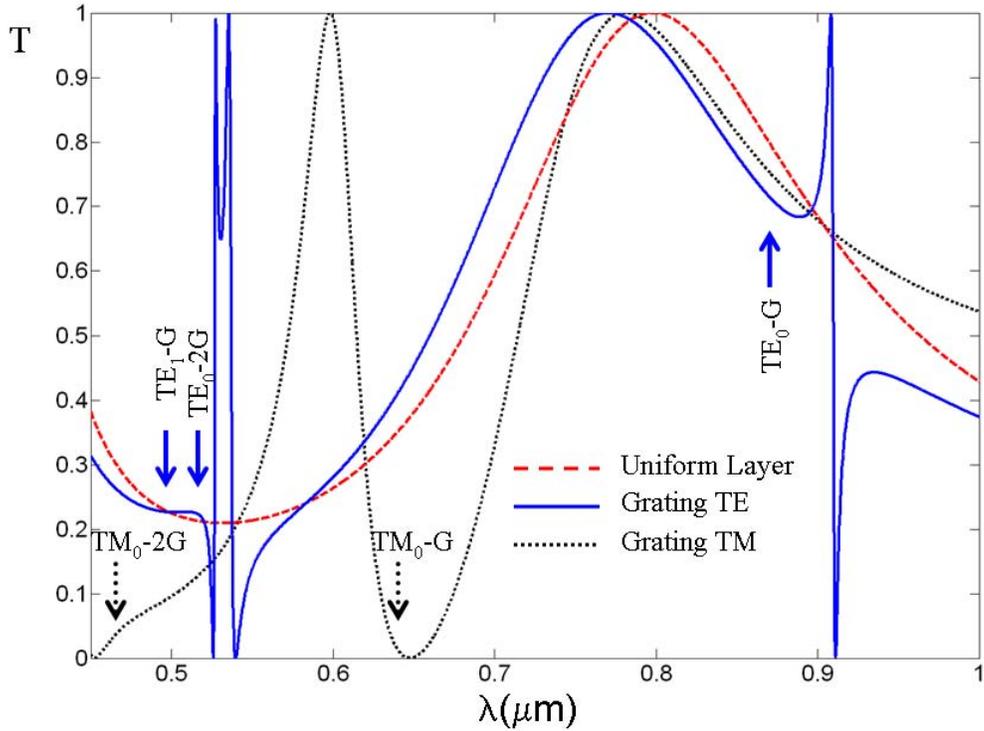

Fig.7. T vs wavelength for TM polarization (dotted line) and TE polarization (continuous line) at normal incidence in the case of a dielectric grating with refractive index n=4.13, thickness d=96nm, period Λ=320nm and slit aperture a=32nm. The dashed line refers to a uniform layer with same thickness and same refractive index. The arrows indicate the wavelengths of the guided modes of the single, uniform layer ($TE_0, TE_1, TM_0, TM_1$) coupled with the reciprocal lattice vectors of the grating. The ridge resonances (guided-mode resonances) are located nearby these wavelengths.

This plot shows that the complex features of the coupling between transverse guided modes and longitudinal Fabry-Perot resonances is not peculiar to surface plasmons, plasmonic structures or TM polarization, but instead it may be obtained in both polarizations in simple dielectric grating geometries. In the figure the arrows indicate the wavelength of the guided modes coupled with the reciprocal lattice vectors of the grating. We observe the transmission resonance for TM polarization (dotted line) located in the valley of the transmission of the Fabry-Perot etalon (dashed line) around λ=532nm and the spectral position of the $TM_0$ guided mode coupled with the reciprocal lattice vector ($TM_0$-G) that is located near the minimum of the transmission resonance, analogous to our previous results for the plasmonic scenario. Moreover we note the two transmission resonances for TE polarization (continuous line) that are located in the valley of the transmission of the Fabry-Perot etalon as well. In the figure we also report the spectral position of the corresponding guided modes $TE_0$ and $TE_1$ coupled respectively with 2G and G. It is evident that the extraordinary transmission for TE polarization [46] could be also explained in terms of GMRs and, more in general, the ridge resonances in the extraordinary transmission scenario, be their nature plasmonic or not, could be analyzed in the framework of the physics of the GMRs [42-45].

## 5. The Rayleigh condition

As we have already seen, in general the plasmon-photon energy-momentum conservation, i.e. Eq. (6.b), corresponds to the transmission minima. In the momentum space, Eq. (6.b) at normal incidence may be written as:

$$k_{SP}(\Omega) = G_m \quad m = 1,2,... \quad . \tag{10}$$

In the $(\Lambda,\lambda)$ plane Eq.(10) can be recast in the following form:

$$\lambda_{SP} = \frac{\Lambda}{m} \quad m = 1,2,... \quad , \tag{11}$$

where $\lambda_{SP}=2\pi/k_{SP}$ is the "SP wavelength" which in our case takes the following simple form: $\lambda_{SP} = \lambda / \mathrm{Re}\sqrt{\varepsilon_{Ag}/(1+\varepsilon_{Ag})}$ with λ the incident wavelength. Eq. (11) tells us that at normal incidence in the $(\Lambda,\lambda)$ plane there are several branches of transmission minima, the first branch occurs when the grating period Λ is *exactly* equal to the SP wavelength $\lambda_{SP}$. Historically, the first explanation of these transmission minima was given by Lord Rayleigh [47], who attributed the transmission minima at normal incidence to the first order diffracted beam becoming grazing to the plane of the grating which occurs when:

$$\lambda = \Lambda. \quad \text{(Rayleigh condition)} \qquad (12)$$

Note that in a photon-plasmon energy-momentum conservation scenario, the Rayleigh condition (12) may be considered as the limiting case of Eq.(11) for m=1 in the low frequency regime, i.e., when the metal tends to behave more similarly to a perfect electric conductor and the SP wavelength tends toward the incident wavelength $\lambda_{SP} \rightarrow \lambda$. So we may expect that, by monitoring the transmission minima of a grating under normal incidence in the $(\Lambda,\lambda)$ plane, the transmission minima will *exactly* follow the law expressed in Eq.(11) and the first branch of transmission minima will only *approximately* be described by the Rayleigh condition and only in the low frequency range. These considerations are indeed confirmed by the calculation shown in Fig. 8.

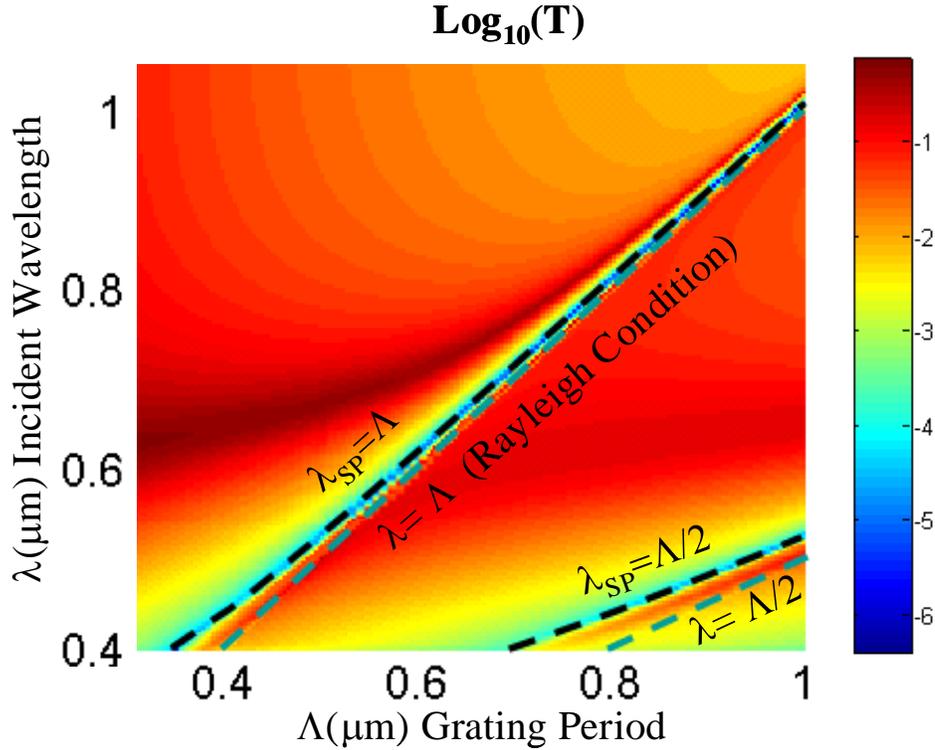

Fig. 8 $Log_{10}(T)$ at normal incidence vs. grating period ($\Lambda$) and incident wavelength ($\lambda$) for a silver grating with thickness d=150nm and slits aperture a=32nm. Note that the first branch of transmission minima follows exactly the plasmonic law, $\lambda_{SP}=\Lambda$, not the Rayleigh condition $\lambda=\Lambda$. Also visible is the second branch $\lambda_{SP}=\Lambda/2$ which departs quite evidently from the line $\lambda=\Lambda/2$.

Our study of the Rayleigh condition (12) vs. its "plasmonic version" (11), shows unambiguously that it is the plasmonic dispersion to rule the transmission minima, consistent with the findings in [7]. The subject has been also analyzed in several publications for 2-D geometries, namely in arrays of sub-wavelength holes, see Ref. [48-49] for example. In the next Section we will analyze the field localization and energy transport in two topical cases, namely: I) the incident field is at normal incidence and it is tuned on the first TM-FP transmission resonance at $\lambda_g/2$; II) the incident field is tuned on the ridge resonance.

## 6. Field Localization and Energy Transport

We start our analysis by showing in Fig. 9 the two transmission resonances that, for exposition purposes, we name respectively with Roman numbers I, II. The structure is the same described in Fig. 4(a). Resonance I is located on the TM-FP($\lambda_g/2$) branch for $\lambda=640$nm and normal incidence with a transmission maximum reaching ~0.6 (or equivalently ~60% of the incoming power). Resonance II is on the ridge at $\lambda=637$nm and $\vartheta=80^0$ (this is the same ridge resonance already represented in Fig. 6). These resonances are characterized by an extremely narrow bandwidth (~1nm), which makes them optimal candidates for a low-power all-optical switching device, for example.

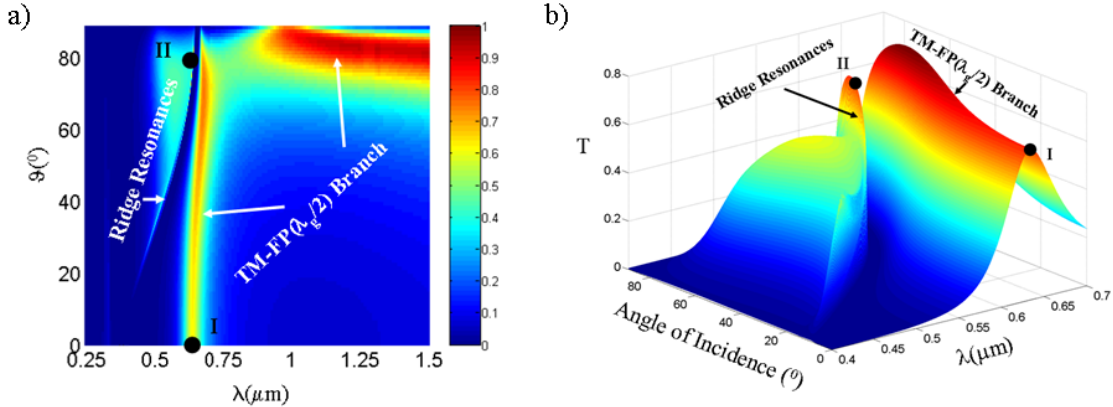

Fig.9. T vs. wavelength ($\lambda$) and incident angle ($\vartheta$) for the case described in Fig.4 (a). a) Topographic view. b) 3-D view from a different perspective. The Roman numbers and the black dots indicate the resonances that we analyze.

In Fig. 10 we show the typical characteristics for the field tuned on resonance I. In particular, Fig. 10(a) reports the square modulus of the magnetic field ($|H|^2$) over the elementary cell. The magnetic field is strongly localized at the center of the slit. This is the typical example of a first-order, single-bell, TM-Fabry-Perot lossy mode that is supported by the grating for a thickness $d \sim \lambda_g/2$. A double-bell-shaped TM-Fabry-Perot mode would resonate at $d \sim \lambda_g$, and so on. In Fig. 10(b) we show the z-component of the Poynting vector. Obviously the energy transport is taking place primarily through the slit, as one may expect, with dramatic energy squeezing through the narrow aperture. In Fig 10(c) we show the sections respectively of $|H|^2$ (continuous line), $|E_x|^2$ (continuous-dotted line), i.e. the square modulus of the the x-component of the electric field, and $S_z$ (dashed line) taken along the x-axis at the centre of the slit (z=75nm). Note that there is a slight penetration of the magnetic field inside the metal, while the electric field remains well-confined in the slit.

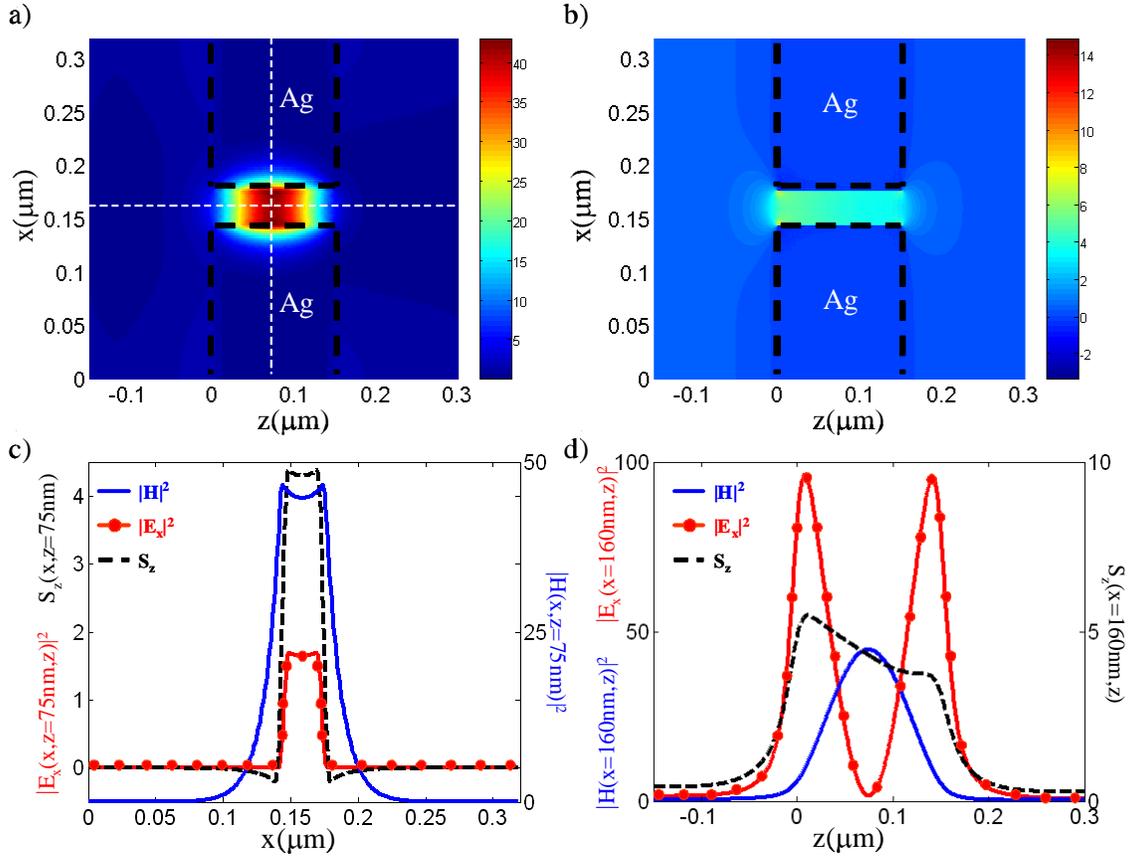

Fig. 10. a) Topographic view of $|H|^2$ vs. x and z in an elementary cell of the grating for the resonance I at λ=640nm and $\vartheta=0^0$. Superimposed (black-dashed line) is the position of the metal. The white lines that cut the figure over the medians represent the sections over which the fields are represented respectively in Fig.10c) (vertical section) and in Fig.10d) (horizontal section). b) Topographic view of $S_z$ vs. x and z. c) Left axis: Sections of $|E_x|^2$ (continuous-dotted line) and $S_z$ (dashed line) along the x-axis, both taken at the centre of the slit (z=75nm). Right axis: Section of $|H|^2$ (continuous line) along the x-axis taken at the centre of the slit (z=75nm). d) Left axis: Sections of $|H|^2$ (continuous line) and $|E_x|^2$ (continuous-dotted line) along the z-axis, both taken at the centre of the slit (x=160nm). Right axis: Section of $S_z$ (dashed line) along the z-axis taken at the centre of the slit (x=160nm).

This penetration of the magnetic field inside the metal must be ascribed to the fact that we are dealing with a metal in a frequency range with poor conductivity, and therefore the mode is not a pure TEM mode, but rather a TM plasmonic mode [24]. In Fig. 10(d) we show a section of the same quantities taken along the z-axis at the center of the slit (x=160nm). Note in Fig. 10(d) the complementary localization of the H and E fields which is typical of a TM-Fabry-Perot mode with antinodes in the E-field at the reflection interfaces of the cavity due to the zero phase change upon reflection at the input and output grating/air interface. Finally in Fig. 11 we report analogous plots for the case of resonance II (the ridge resonance of Fig. 5) at λ=637nm and ϑ=80$^0$.

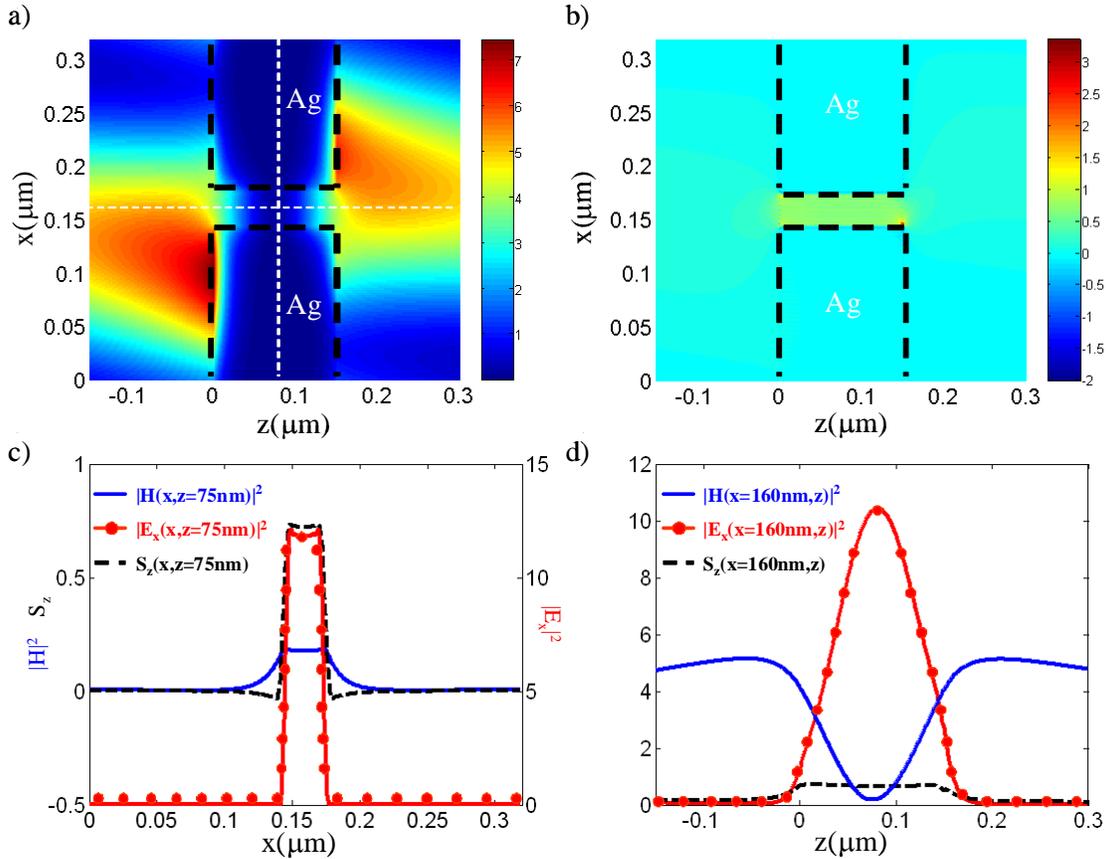

Fig.11. a) Same as in the caption of Fig 10a) for the resonance II at λ=637nm and ϑ=80$^0$. b) Topographic view of $S_z$ vs. x and z. c) Left axis: Sections of $|H|^2$ (continuous line) and $S_z$ (dashed line) along the x-axis, both taken at the centre of the slit (z=75nm). Right axis: Section of $|E_x|^2$ (continuous-dotted line) along the x-axis taken at the centre of the slit (z=75nm). d) Sections of $|H|^2$ (continuous line), $|E_x|^2$ (continuous-dotted line) and $S_z$ (dashed line) along the z-axis, all three taken at the centre of the slit (x=160nm).

Comparing the field localizations characteristic of resonance I and resonance II, one immediately realizes that they are almost complementary. The magnetic field is localized inside the slit for resonance I, while it is localized outside the slit for resonance IV (compare Fig.10(a) and Fig.11(a)). In particular from Fig. 11(d) we see that this time is the electric field to be strongly localized at the centre of the slit, with a single-bell-shaped envelope, in contrast to the case of Fig.10(d). This complementary behavior of the field localization is, again, reminiscent, for example, of the band-edge resonances of finite 1-D PC [41] where the electric field can be localized over the high index layers and the magnetic field over the low index layers or vice-versa depending on whether the field is tuned at the high frequency or low frequency band-edge. Regarding the energy transport, we note that the flux of energy inside the slit along the z-direction is greater for resonance I ($S_z$~4) than for resonance II ($S_z$~0.7), although for both resonances the transmission is approximately the same (0.6 or 60%). This fact should not be surprising if we take into account that the incoming field for resonance I is at normal incidence, while for resonance II is at an angle of $80^0$, therefore the incoming power along the z-direction for resonance II is rescaled by a factor $\cos(80^0)$ with respect to the incoming power for resonance I. Let us finally briefly comment that it could be very intriguing to study the optical bistability of these ridge-resonances by filling the slit with a Kerr medium, as we plan to do in the near future. In fact, these resonances seem to show all the characteristics necessary to achieve a low-power all-optical switching: they are very narrow, the electric field is highly localized in the slit, and moreover, given their particular ridge-like nature, the switch could take place both for positive and negative values of the $\chi^{(3)}$ coefficient.

## 7. Conclusions

In conclusion, we have highlighted here some interesting properties of the transmission resonances of sub-wavelength metallic gratings with extremely narrow slits under TM light polarization, elucidating the interplay between the TM-FP branch and the SP dispersion. It is in place here a final note regarding the role of SPs in the enhanced transmission from metallic gratings at lower frequencies (far-IR, THz, microwave) [36]. In the low frequency regime the metal becomes closer and closer to a perfect conductor and the SPs wavelength becomes closer and closer to the incident wavelength. Therefore we expect results qualitatively similar to those presented here, but with few caveats: a) the peak of transmission of the FP branch will be much closer to the 100% transmission due to less dissipation from the lateral surfaces of the slits during the guiding; b) the interplay between the FP branch and the SP dispersion still continues to be ruled by Eqs.(6), but the SP momentum $k_{SP}$ becomes practically almost indistinguishable from the normal incidence momentum of the incoming photon ($k_{SP} \rightarrow k_0$) due to the rectification of the SP dispersion; c) the guided quasi-TEM mode will have an effective index close to 1 ($n_{eff}\sim 1$) so that the positions of the FP transmission resonances will be ruled approximately by the incident wavelength λ [27-28]. We believe that the results we have presented here may improve the understanding of enhanced transmission through 1-D plasmonic or metallic gratings at optical frequencies.

**Acknowledgments**

We thank M. Scalora, N. Akozbek and M. Buncick for helpful discussions. The work of G.D. and N.M has been partially supported by DARPA projects "Nonlinear Plasmonic Devices" and "Tunable Metamaterials". A. A. was partially supported by AEgis Technologies Group, Inc., with contract #41-STTR-UTX-0652.